# Fiscal shocks and asymmetric effects: a comparative analysis

**Ioannis Praggidis, Periklis Gogas, Vasilios Plakandaras, Theophilos Papadimitriou**

Department of Economics

Democritus University of Thrace

Komotini, 69100, Greece

## Abstract

We empirically test the effects of unanticipated fiscal policy shocks on the growth rate and the cyclical component of real private output and reveal different types of asymmetries in fiscal policy implementation. The data used are quarterly U.S. observations over the period 1967:1 to 2011:4. In doing so, we use both a vector autoregressive and the novel support vector machines systems in order to extract the fiscal policy shocks series. The latter has never been used before in a similar macroeconomic setting. Within our research framework, in order to test the robustness of our results to alternative aggregate money supply definitions we use two alternative moentary aggregates. These are the commonly reported by central banks and policy makers simple sum monetary aggregates at the MZM level of aggregation and the alternative CFS Divisia MZM aggregate. From each of these four systems we extracted four types of shocks: a negative and a positive government spending shock and a negative and a positive government revenue shock. These eight different types of unanticipated fiscal policy shocks are next used to empirically examine their effects on the growth rate and the cyclical component of real private GNP in two sets of regressions: one that assumes only contemporaneous effects of the shocks on output and one that is augmented with four lags of each fiscal shock.

**Acknowledgement**: We would like to thank one anonymous referee for his comments that greatly improved our manuscript.

# 1. Introduction

In this paper we empirically test the existence of non-linearities that may be associated with the conduct of fiscal policy. In doing so, we try to detect two types of fiscal policy asymmetries: first, whether equal in magnitude contractionary or expansionary fiscal shocks have the same multiplier impact on real output, and second whether theoretically equal –in terms of their impact on the government budget fiscal policy tools, such as a tax cut or an increase in government spending, have the same impact on output.

Fiscal and monetary policies are the cornerstone of policymaking. However, until 2000 the main bulk of empirical research was dedicated solely to the effects of monetary policy. In the aftermath of the global crisis of 2008 there is a growing debate of whether governments should run fiscal stimulus packages in order to restore previous growth rates or run an austerity program to reduce deficits and in the long-run debt as a percent of GDP. Recently for example, highly indebted Eurozone countries (Greece, Ireland, Portugal and Spain) are required to implement fiscal austerity measures in order to balance their balance sheets. In this context it is interesting to see whether and how Keynesian principles may apply.

According to (Bertola and Drazen, 1993), governments should choose fiscal stimulus packages if they accept a positive and above unity fiscal multiplier regardless of the debt to GDP ratio. Keynesian economics assert that government spending and tax cuts, directly affect disposable private income and through the channel of active demand the economy tracks itself back to a growth path. The fiscal multiplier under Keynesian beliefs is well above unity as there is no crowding out effect and the wealth effect is not so strong. Due to various rigidities in the markets (labor, goods and services), this fiscal stimulus during recessions and fiscal contraction during boom times accordingly, is necessary and appropriate in order to restore equilibrium. Although, the exact value of the multiplier depends on various other factors, such as the simultaneous usage of monetary policy, the openness of the economy, the exchange rate regime e.t.c. its sign, however, is not under question: we expect a positive impact on GDP from an increase in government spending.

The neoclassical school on the other hand, asserts that government spending or tax cuts have no impact on GDP due to the Ricardian equivalence (Barro 1974). Agents fully anticipate the debt burden of the fiscal stimulus, expecting higher taxes in the future (wealth effect). Thus, in order to smooth out their level of consumption they save more as their disposable income increases leaving private consumption unchanged.. There is a crowding out effect of the private sector that fully offsets the increase of the demand from the public sector which renders the fiscal multiplier to zero. This is more apparent in periods of growth, since then the probability of a more efficient usage of resources from the government is lower than it is during a recession. On the other hand, there is room for a low positive multiplier during recessions, since resources are underused.

There is also a new class of research pointing to an exactly different direction than that of Keynesian economics: these find that the multiplier of fiscal contraction is positive and vice versa. This is known as contractionary fiscal expansion effect or expansionary fiscal contraction due mostly to a wealth effect that is, consumers put more weight to future consumption than to current one. At this notion, Alesina and Perotti (1997) and Giavazzi and Pagano (1990) among others, state that fiscal contraction based on expenditure cuts may be expansionary if it is accompanied by currency devaluation or by agreements with the unions. The greater this adjustment is the more is being anticipated by the agents leading to more powerful results. Furthermore, a tax increase in order to accommodate a deficit has the exact opposite results than a decrease of government spending because it reduces the competitiveness of the economy. This view is enhanced by Blanchard (1990), who states that fiscal consolidation may reduce uncertainty for the future leading to an increase in household's wealth today. This can be achieved through the decrease of interest rates as a result of the reduction of the risk premium of government bonds (Alesina and Ardagna, 2009).

In their seminal paper, Bertola and Drazen (1993), postulate that the sign of the fiscal multiplier depends on the GDP to debt ratio. In a hypothetical economy, where all agents are rational, and GDP to debt ratio is low, an increase of the government spending will be neutral to the real economy, featuring a Ricardian or even a negative effect. If the GDP to debt ratio is relatively large a fiscal consolidation signals a trial of the government to stabilize the economy and thus lifting future uncertainty leading to a positive multiplier or to an anti – Keynesian effect.

According to the above, the fiscal multiplier for an increase in government expenditures can range between negative and positive values and be large or small. According to the above, we can identify five potential sources of non – linearities/asymmetries of fiscal policy: a) the phase of the business cycle, b) the GDP to debt ratio, c) the sign of the shock (positive versus negative shocks of the same instrument), d) the nature of the shock (spending versus revenues), e) the magnitude of the shock.

In this paper, we try to estimate the value of the fiscal multiplier taking into account the sign and the nature of the shock. Using VAR analysis with identified structural errors, Machine Learning (ML) techniques, a new dataset for the U.S. economy and running various tests, we come along some very interesting results. We cannot reject asymmetries in government spending between a positive unanticipated government spending shock and a negative such shock. The same asymmetries are detected for the unanticipated government revenue shocks. We also detect asymmetries in expansionary and contractionary unexpected fiscal policies.

The remainder of this paper is organized as follows: Section 2 presents the empirical literature review, Section 3 provides a detailed description of the data and the methodology used. The main results are presented in Section 4 and Section 5 concludes.

## 2. Literature Review

Despite this divergence of opinions, the empirical research is too narrow and is divided between linear and nonlinear policy analysis. Linear analysis covers most of the research, while nonlinear analysis is being implemented only in recent years. Empirical research focused into fiscal policy in the last decade following mostly the seminal work of Blanchard and Perotti (2002) in a VAR analysis which was built upon the innovative work of Sims (1980) in VAR analysis. Blanchard and Perotti (2002) introduced a new method of identification of structural errors using institutional information on tax and transfer system and under the main assumption, among others, that fiscal policy is a rather long process using quarterly data introduce their restrictions and identify structural fiscal shocks that are exogenous to the rest of the VAR variables. They conclude that, the U.S. economy experiences Keynesian effects regarding the sign of fiscal multipliers as well as there are asymmetries between tax and government purchases multipliers but not asymmetries of the effects on the output of a positive versus a negative change in taxes. Tagkalakis (2008) using an unbalanced yearly panel data set (1970-2002) of nineteen OECD countries, confirmed that in the presence of binding liquidity constraints during recessions both positive government spending and negative tax shocks have stronger stimulus effects on private consumption than in expansions. In a different analytical framework Leeper et al. (2010) show that government investment is contractionary in the short run, at worst, and has a muted impact, at best. This is mainly due to substantial time to build lags. The results over the long run are conditional upon the productivity of the public capital. Pereira and Lopes (2010) examining U.S. quarterly data over the 1965:2 to 2009:2 period in a Blanchard-Perron identification mode into a Bayesian simulation procedure, they find that policy effectiveness has come down substantially. More specifically, this trend is more evident for taxes net of transfers than for government expenditures, although, fiscal multipliers keep Keynesian signs. Cogan et al. (2009), focusing on an empirically estimated macroeconomic model for the U.S., find that the government spending multipliers are much less in new Keynesian that in old Keynesian models. The multipliers are less than one as consumption and investment are crowded out. On the other hand, Romer and Romer (2010), using new sources of data such as presidential speeches, executive-branch documents and Congressional reports, identify the size, timing and principal motivation for all major post-war tax policy actions. Their main findings indicate a very large effect of tax changes on output and on investments. This multiplier is well above unity, being in stark contrast with the findings of previous empirical researches. Barro and Redlick (2009), estimate a multiplier regarding responses of U.S. GDP to changes in defence spending between 0.6-0.7. As they point out in their paper, the exact volume of the multiplier is subject to economic slack, reaching unity as unemployment rate is quite high, around 12%. Positive tax rate shocks have significantly negative effects on real GDP growth. Mountford and Uhlig (2009), incorporating a VAR analysis and using new restrictions to identify revenue and spending shocks, as well as taking into account business cycle and monetary shocks, conclude that deficit financed tax cuts are the best fiscal policy to improve GDP, finding a very large multiplier. Gali et al. (2007), show that in an economy in which for some households (named rule of thumb consumers) consumption equals labor income and there exist sticky prices, it is possible that government spending shocks positively affect consumption. In this way, wealth effects are totally overshadowed by the sensitivity to current disposable income. Aggregate demand is partly insulated from the negative wealth effect generated by the higher levels of taxes needed to finance the fiscal expansion.

In a non-linear framework, Baum and Koester (2011), using a threshold VAR model, analyse the effects of fiscal policy on economic activity over the business cycle for Germany. They derive a fiscal multiplier around 0.7 for both revenues and spending in a linear model. When they take into account the phase of the business cycle, they find a spending multiplier around unity in boom times and 0.36 in recessions. There are also non linearities regarding the sign of government intervention through spending. With respect to revenue shocks they find less diverging results for both the phase of the business cycle and the type of fiscal policy implemented (expansionary or contractionary).

As it is clear from the above, empirical research spans a wide range of tests, including linear and non–linear models concerning the phase of the business cycle, the financial constraint of the agents, the nature and the sign of the fiscal intervention. Most of these studies, converge to fiscal multiplies below unity with the spending multiplier being of greater importance than the tax multiplier. In what follows we try to unfold the impact of fiscal policy using quarterly data for the U.S. economy for government spending, total government revenue, GNP (growth rate and cyclical component) and monetary variables such as the Treasury bill rate and the money supply. In this study we introduce four main innovations: first, to the best of our knowledge the Divisia monetary aggregates have not yet been used to previous research pertaining to fiscal policy. Second, following Cover's (1992) procedure of identifying monetary policy shocks we extract the unanticipated fiscal policy shocks on government spending and revenue. Moreover, we introduce Support Vector Regression models which have never been used before in this research area, but the empirical results of other implementations such as exchange rate (Papadimitriou et al, 2013), bank insolvency (Papadimitriou et al, 2013) and GDP forecasting (Gogas et al, 2013) imply high generalization abilities with non-linear and non-stationary datasets. Finally, we explicitly test for the asymmetric effects on the growth rate and the cyclical component of real private GNP of a contractionary and expansionary fiscal policy. We come up with three key findings; first, all fiscal multipliers are below unity but with signs as predicted by Keynesian theory. Second, government expenditures have a larger impact as compared to the tax policy and finally, positive government spending shocks are more significant than negative spending shocks. All these results are in line with previous studies and are robust through many tests using structural identification proposed by Blanchard and Perotti (2002).

## 3. The Data

In this study we use quarterly data that span the period 1967Q1 to 2011Q4. The range of the data sample is limited by the availability of the monetary aggregates. The data are taken from the St. Louis Federal Reserve Economic Data (FRED) service. These include the real private Gross National Product, government consumption expenditures and gross investment, government current receipts and the 3-month Treasury bill rate[1]. All initial data are in current values and they are transformed –with the exception of the Treasury-bill rate- to real series by using the implicit price deflator of the GNP with 2005 as the base year. The two monetary aggregates used in this study are the official simple-sum aggregates in the MZM level of aggregation as they are reported by the Federal Reserve Bank of St. Louis and the Divisia MZM aggregates both in real terms. The Divisia monetary aggregate series are from the new Divisia monetary aggregates maintained within the Center of Financial Stability (CFS) program Advances in Monetary and Financial Measurement (AMFM), called CFS Divisia aggregates and documented in Barnett et al. (2013). We use both types of monetary aggregates in an effort to see whether our results are affected by the so-called "Barnett critique". In this regard, Barnett (1980) argues that official simple-sum monetary aggregates, constructed by the Federal Reserve, produce an internal inconsistency between the implicit aggregation theory and the theory relevant to the models and policy within which the resulting data are nested and used. That incoherence has been called the Barnett Critique [see, for example, Chrystal and MacDonald (1994) and Belongia and Ireland (2013)], with emphasis on the resulting inference and policy errors and the induced appearances of function instability. Finally, all data with the exception of the Treasury-bill rate are transformed to natural logarithms. To test the integration properties if our data we perform three different unit root tests: a) an augmented Dickey-Fuller test, b) a KPSS test where the null hypothesis is stationarity and finally c) an Elliott-Rothenberg-Stock test. In Table 1 we present the results of these unit root tests and we conclude that all variables used in this study are I(1). Thus for the rest of the empirical section we use the first differences of the variables unless otherwise stated.

---

[1] The relevant FRED codes are GNPC96, GCEC, GRECPT and TB3MS respectively.

## 4. Empirical Model and Identification

Since we determined in the previous section that all our variables are I(1) we proceed by testing our variables for a commons stochastic trend. Table 3, reports the results of the Johansen maximum likelihood cointegration tests on a VAR with lag length $p = 3$. We also report tail areas of residual misspecification tests. Two test statistics are used to test for the number of cointegrating vectors, the trace ($\lambda_{trace}$) and maximum eigenvalue ($\lambda_{max}$) test statistics. In the trace test the null hypothesis that there are at most $r$ cointegrating vectors is tested against a general alternative. In the case of the maximum eigenvalue test the alternative is explicitly stated. Using 99% critical values for the two tests we that the $\lambda_{trace}$ and $\lambda_{max}$ test statistics provide evidence of one cointegrating relation in both VAR models we test: one with the simple sum MZM as the exogenous monetary aggregate and the other with the CFS Divisia instead. Since we detected one cointegrating vector, we proceed in our analysis by using a Vector Error Correction Model (VECM) including the lagged error correction term in the VARs regressors.

As it was previously mentioned, we use a structural VAR model with two alternative sets of exogenous variables. We perform a Blanchard-Perotti (2002) identification procedure to extract the structural errors. The basic reduced form VAR specification in order to identify the structural errors is:

$$Y_t = A_0 + \Gamma ec_{t-1} + A(L,q)Y_{t-1} + BZ_t + U_t, \tag{1}$$

where $Y_t$ is a three dimensional vector of the endogenous variables, government revenue ($r$), government spending ($g$) and real private Gross National Product ($y$) and $Z_t$ is a vector of the two exogenous monetary variables, the 3-month Treasury bill rate (TB3) and the monetary aggregate, where we use alternatively a simple-sum and a CFS Divisia in the MZM level of aggregation. All variables are in first differences. The exogenous variables vector is two dimensional because we use each monetary aggregate separately along with the TB3 variable. $U_t$ represents the three dimensional vector of reduced form residuals with the corresponding ordering $[r_t, g_t, y_t]$ and finally $A_0$ is the intercept coefficient vector, $A(L_q)$ is a four lag polynomial and $B$ is the exogenous variables coefficients vector. A four quarter lag length is chosen as there is a seasonality pattern in the response of taxes to output – see Blanchard-Perotti (2002).

### 4.1 Blanchard-Perotti Identification

We employ the Blanchard-Perotti (2002) method of structural identification. As they well document in their seminal paper, the innovations in the fiscal variables, taxes and revenues are a linear combination of three types of shocks, a) the automatic response of these fiscal variables to output (automatic stabilizers), b) the discretionary effects of revenues to spending shocks and vice versa, c) the random fiscal shocks which are to be identified. Thus, the equation system is:

$$t_t = a_1 \varepsilon_t^{GNP} + a_2 \varepsilon_t^G + \varepsilon_t^T$$

$$g_t = b_1 \varepsilon_t^{GNP} + b_2 \varepsilon_t^T + \varepsilon_t^G$$

$$GNP_t = c_1 \varepsilon_t^T + c_2 \varepsilon_t^G + \varepsilon_t^{GNP}$$

In order to set the appropriate restrictions Blanchard and Perotti further assume that the first set of shocks ($a_1$ and $b_1$ for taxes and spending respectively) can be estimated as the elasticity of fiscal variables to output shocks as it takes more than a quarter for a fiscal policy measure to be decided and be implemented. As a measure for tax elasticity on output, $a_1$, we take into consideration Blanchard-Perotti's calculations who report an average value of 2. As for the spending multiplier, $b_1$, this is set to zero, as the main component of primary government spending, unemployment transfers is included in net revenues[2]. Then, contemporaneous effect of fiscal variables to output ($c_1$ and $c_2$) need to be estimated. Again in line with Blanchard et al (2002) and Baum et al. (2011), we use the cyclically adjusted reduced form fiscal policy shocks and we estimate the third equation of the equation system 2. Finally, under the assumption that revenue decisions come first, $a_2$ is set to zero. This is so, because $a_2$ represents the discretionary response of revenues to spending.

### 4.2 Support Vector Regression

The Support Vector Regression is a direct extension of the classic Support Vector Machine algorithm proposed by Vladimir Vapnik (1992). When it comes to SVR, the basic idea is to find a linear function that has at most a predetermined deviation from the actual values of the actual dataset. In other words we do not care about the error of each forecast as long as it doesn't violate the threshold, but we will not tolerate a higher deviation. The Support Vector (SV) set which bounds this "error-tolerance band" is located in the dataset through a minimization procedure.

One of the main advantages of the SVR in comparison to other machine learning techniques is the ability to identify global minima avoiding local ones, thus reaching an optimal solution. This aspect is crucial to the generalization ability of the SVR model in producing accurate and reliable forecasts. The model is built in two steps: the training step and the testing step. In the training step, the largest part of the dataset is used for the estimation of the function (i.e. the detection of the Support Vectors that define the band); in the testing step, the generalization ability of the model is evaluated by checking the model's performance in the small subset that was left aside in the first step performing out-of-sample forecasting.

Using mathematical notation and starting from a training dataset $D = [(\mathbf{x}_1, y_1), (\mathbf{x}_2, y_2), \dots (\mathbf{x}_n, y_n)], \mathbf{x}_i \in R^m, y_i \in R, i = 1,2, \dots n$, where for each dataset pair, $\mathbf{x}_i$ are observation samples and $y_i$ is the dependent variable (the target of the regression system that we need to approximate). As the scope is to minimize the loss function $\min\left(\left(\frac{1}{2}\|\mathbf{w}\|^2\right)\right.$ subject to $\left|y_i - (\mathbf{w}^T\mathbf{x}_i + b)\right| \leq \epsilon, i = 1,2, \dots, n$, we enforce an upper deviation threshold and creating an "error-tolerance" band. Expanding this initial framework, Vapnik and Cortes (1995) proposed a soft margin model, i.e. accept the existence of vectors outside this error tolerance zone that are penalized according to their distance from the zone. So, they introduce slack variables to the loss function $\zeta_i$ and $\zeta_i^*$ controlled through a cost parameter C, resulting to a loss function $\min\left(\frac{1}{2}\|\mathbf{w}\|^2 + C\sum_{i=1}^{n}(\zeta_i + \zeta_i^*)\right)$. In this way the primal problem that we wish to minimize is:

$$L := \frac{1}{2}\|\mathbf{w}\|^2 + C\sum_{i=1}^{n}(\zeta_i + \zeta_i^*) - \sum_{i=1}^{n}(\eta_i\zeta_i + \eta_i^*\zeta_i^*) - \sum_{i=1}^{n} a_i(\varepsilon + \zeta_i - y_i + \mathbf{w}^T\mathbf{x_i} + b)$$
$$- \sum_{i=1}^{n} a_i^*(\varepsilon + \zeta_i^* + y_i - \mathbf{w}^T\mathbf{x_i} - b) \qquad (2)$$

where $\eta_i, \eta_i^* a_j, a_j^*$ are the Lagrange multipliers from the Lagrangian primal function (2).

---
[2] For an extended presentation see Blanchard et al (2002) and Baum et al. (2011).

Instead of solving (1) we attack the dual form of the problem, which takes the form:

$$\max\left(\frac{1}{2}\sum_{i,j=1}^{n}(a_i - a_i^*)(a_j - a_i^*)\mathbf{x_i}^T\mathbf{x_j} - \varepsilon\sum_{i=1}^{n}(a_i - a_i^*) + \sum_{i=1}^{n} y_i(a_i - a_i^*)\right) \qquad (3)$$

Subject to $\sum_{i=1}^{n}(a_i - a_i^*) = 0$ and $a_i, a_j \in [0, C], \quad i, j = 1,2,3,\ldots, n$

The solution of the primal problem (2) is

$$\mathbf{w} = \sum_{i=1}^{n}(a_i - a_i^*)\mathbf{x_i} \qquad (4)$$

and
$$y = \sum_{i=1}^{n}(a_i - a_i^*)\mathbf{x_i}^T\mathbf{x} \qquad (5)$$

Real life phenomena, are rarely modelled by linear functions accurately. A simple way to bypass the problem is by projecting the observed dataset from the initial data space into a higher dimensional space were the linear model is appropriate. The "kernel trick" follows the projection idea while ensuring minimum computational cost: the dataset is mapped in an inner product space, where the projection is performed using only dot products (through special "kernel" functions) within the original space, instead of explicitly computing the mapping of each data point. The SV methodology coupled with the "kernel trick" is a very powerful tool for classification and regression. Non-linear kernel functions has evolved the SVR mechanism to a non-linear regression model, able of approximating non-linear phenomena.

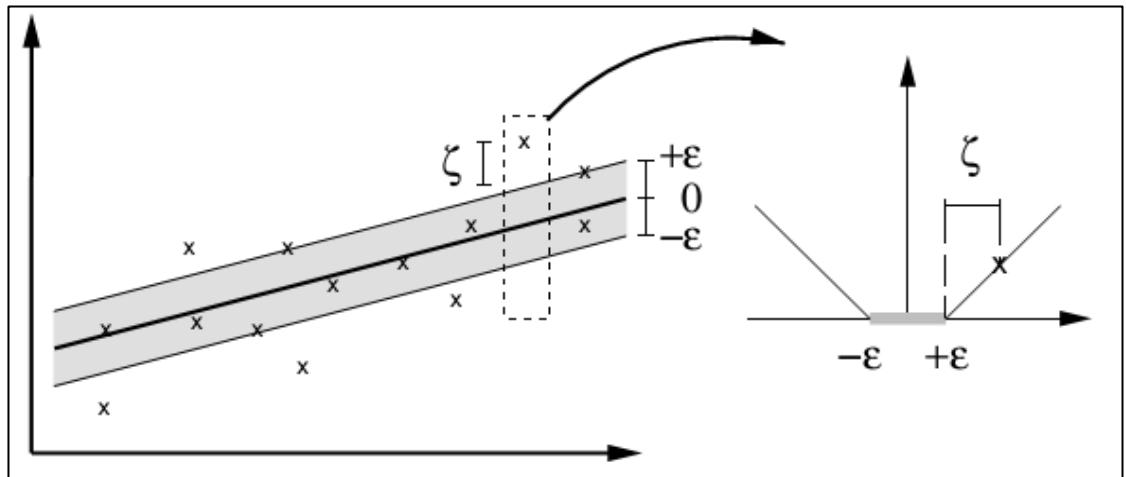

**Figure 1**: Upper and lower threshold on error tolerance indicated with letter ε. The boundaries of the error tolerance band are defined by Support Vectors (SVs). On the right we see the projection form 2 to 3 dimensions space and the projected error tolerance band. Forecasted values greater than ε get a penalty ζ according to their distance from the tolerance accepted band (source Scholckopf and Smola, 1998).

The described mapping was performed on four kernels: the linear, the radial basis function (RBF), the sigmoid and the polynomial. The mathematical representation of each kernel is:

Linear $$K_1(\mathbf{x}_1, \mathbf{x}_2) = \mathbf{x}_1^T \mathbf{x}_2 \tag{6}$$

RBF $$K_2(\mathbf{x}_1, \mathbf{x}_2) = e^{-\gamma \|\mathbf{x}_1 - \mathbf{x}_2\|^2} \tag{7}$$

Polynomial $$K_3(\mathbf{x_1}, \mathbf{x_2}) = \left(\gamma \mathbf{x}_1^T \mathbf{x}_2 + r\right)^d \tag{8}$$

Sigmoid(MLP) $$K_4(\mathbf{x}_1, \mathbf{x}_2) = \tanh\left(\gamma \mathbf{x}_1^T \mathbf{x}_2 + r\right) \tag{9}$$

with factors d, r, γ representing kernel parameters.

Again the procedure was repeated with SVR and we extract both fiscal policy structural errors: from government revenue and the government spending vector. Two sets of such errors are used: the ones from the simple sum MZM monetary aggregate as an exogenous variable VAR and the ones from the VAR with the Divisia MZM as an exogenous variable.

## 5. The Empirical results

Following Cover (1992), from each of the above two systems (MZM and CFS) we extract the residual series from the equations of government revenue (r) and government spending (g). These represent the unanticipated fiscal policy shocks. The series of the negative government spending shocks equals the government spending shock if the latter is negative otherwise it is equal to zero. The series of the positive government spending shocks equals the government spending shock if this is positive and otherwise it is equal to zero. In the same manner we construct the negative and positive government revenue shocks. Formally:

$$SGN_t = -1/2\,[|GSS_t| - GSS_t]$$

$$SGP_t = 1/2\,[|GSS_t| + GSS_t],$$

where $GSS_t$ is the government spending shock extracted as described above. In a similar manner we construct the negative and positive government revenue shocks $SRN_t$ and $SRP_t$.

## 5.1 Systems with contemporaneous shocks

In the previous section we extracted four series of unanticipated fiscal policy shocks, from each one of the two VARs considered in this study as well as for the support vector regression model. For each VAR these are the negative and positive government spending shocks and the negative and positive government revenue shocks series: $SGP_t$, $SGN_t$, $SRP_t$ and $SRN_t$ respectively. In order to investigate the possible existence of fiscal asymmetries, following Cover (1992), we run the following regression with each of the two sets of unanticipated fiscal policy shocks:

$$d(y_t) = a_0 + \sum_{i=1}^{2} a_i d(y_{t-i}) + \gamma\, d(tb3_t) + \delta d(m_t) + \beta_1 SGP_t + \beta_2 SGN_t + \beta_3 SRP_t +$$
$$\beta_4 SRN_t + e_t, \tag{11}$$

where $d(y_t)$ is the first difference of the real private GNP at period $t$, $d(y_{t-i})$ are two lags of the output first differences, $SGP_t$, $SGN_t$, $SRP_t$ and $SRN_t$ are the extracted unanticipated fiscal shocks to the economy as discussed above and $a_0, a_i, \gamma, \delta$, and $\beta_i$ are parameters to be estimated. In these systems we assume that only current fiscal policy shocks affect the real output growth level and thus we include no lagged values of the fiscal shocks. We also estimate equation (11) with the cyclical component of the real private GNP as the dependent variable. The cyclical component of real private GNP is extracted using a standard Hodrick-Prescott filter with $\lambda = 1600$. The empirical results are presented in Table 4. Systems 1 and 3 present the results of the estimation of equation (11) with the fiscal policy shocks identified from the VAR including the simple sum MZM monetary aggregate and Divisia MZM monetary aggregate respectively. System 5 includes the estimates of equation (11) when the dependent variable is the cyclical component of real private GNP and the fiscal policy shocks are identified from a VAR with the Divisia MZM as the monetary aggregate variable[3]. In system ML we present the results of the estimation of equation (11) with the fiscal policy shocks identified from the support vector regression model.

The estimated coefficients and the reported *p*-values of the fiscal policy shocks provide evidence on the significance and magnitude of the multipliers of the various fiscal shocks on the growth rate of real private GNP and its cyclical component. Moreover, at the lower part of Table 4 we report the tail areas of the *F*-tests performed in testing for fiscal policy asymmetries. First, we test the null hypothesis that the multiplier of a positive government spending shock is equal to the multiplier of a negative government spending shock ($H_0: \beta_1 = \beta_2$) or in other words that a contractionary government spending shock has a symmetric effect on output as an equal expansionary government spending shock. Second, in a similar manner, we test for symmetric effects of the contractionary and expansionary government revenue shocks ($H_0: \beta_3 = \beta_4$). Next, we try to investigate whether equivalent in terms of their impact on government deficit fiscal policies have symmetric effects on the level and growth rate of real GNP. First, we test policies that increase the deficit, positive government spending and negative government revenue shocks ($H_0: \beta_1 = \beta_4$) and finally shocks that lead to fiscal consolidation, a decrease in government spending and an increase in government revenue shock ($H_0: \beta_2 = \beta_3$).

According to Table 4 we have some interesting results. First, all coefficients have the signs expected by theory in all systems except for the system ML. The coefficients of positive government spending shocks (SGP) have positive signs as they are expected to increase real private GNP. The same sign is expected on the coefficients of the negative government spending shocks (SGN): a negative shock multiplied by a positive coefficient produces a decrease in real private GNP. For analogous reasons the estimated coefficients on unanticipated revenue shocks have negative signs: an unexpected increase in government revenue (SRP) is expected to reduce real private GNP and an unexpected decrease in government revenue (SRN) will increase real private GNP.

---

[3] The relevant estimates with identified fiscal policy shocks from the VAR with the simple sum MZM monetary aggregate as the exogenous variable are not included here as they are qualitatively exactly the same as the ones of system 5. They are available of course from the authors upon request.

We detect some asymmetries across all three systems with respect to fiscal policy shocks. It appears that in the System 1 and 2 where the dependent variable is the growth rate of real private GNP expansionary and contractionary unanticipated fiscal policies have asymmetric effects: the expansionary fiscal policy through either a positive unanticipated government spending shock (SGP) or a negative unanticipated government revenue shock (SRN) is statistically significant with p-values of 0.008 and 0.052 respectively. On the contrary, a contractionary fiscal policy through either a negative unanticipated government spending shock (SGN) or a positive unanticipated government revenue shock (SRP) is not significant even at the 0.10 significance level. In System 5 where the dependent variable is the cyclical component of real private GNP it appears that only a positive unanticipated government revenue shock has some impact on the cycle. Both government spending and revenue tests show that in general expansionary unanticipated fiscal policy shocks have asymmetric effects and appear to affect real private output more than contractionary fiscal shocks as the later appear insignificant. Expansionary fiscal policy is significant either through spending or revenue. Nonetheless, the coefficient of the positive unanticipated government spending shock is more than three times larger than the coefficient of the negative unanticipated government revenue shock in Systems 1 and 3 and more than two times larger when the dependent variable is the cyclical component of real private GNP. In the lower part of Table 4, the F-tests show that this asymmetry is statistically significant only for System 1. In system ML the unanticipated fiscal policy shocks does not seem to have any statistically significant impact on the level of the real private GNP.

## 5.2 Systems augmented with lagged shocks

In this section we augment the regressions run in equation (11) by assuming that not only the current values of the explanatory variables affect the level and growth rate of GDP but also four lags that correspond to a year's worth of historical information. The estimated equation now becomes:

$$d(y_{t)} = a_0 + \sum_{i=1}^{4} a_i\, d(y_{t-i}) + \sum_{i=0}^{4} (\gamma_i\, tb3_{t-i} + \gamma_i\, d(m_{t-i}) + \beta_{1i}\, SGP_{t-i} + \beta_{2i}\, SGN_{t-i} + \beta_{3i}\, SRP_{t-i} + \beta_{4i}\, SRN_{t-i}) + e_t \qquad (12)$$

with similar specification as equation (11). The results from running equation (12) are presented in Table 5. Systems 2 and 4 have the growth rate of real private GNP as the dependent variable with fiscal policy shocks identified from a VAR with simple sum and Divisia MZM as the monetary aggregate exogenous variable respectively. System 6 is estimated with the cyclical component of real private GNP as the dependent variable and the identified residuals from the VAR with the Divisia MZM monetary aggregate[4]. System ML is estimated with the growth rate of real private GNP as the dependent variable with fiscal policy shocks identified from a support vector regression model.

The estimated coefficients and the reported *p*-values of the contemporaneous and lagged fiscal policy shocks provide evidence on the significance and magnitude of the multipliers of the various fiscal shocks on the level and growth rate of real GNP. In the lower part of Table

---

[4] Again, the relevant estimates with identified fiscal policy shocks from the VAR with the simple sum MZM monetary aggregate as the exogenous variable are not included here as they are qualitatively exactly the same as the ones of system 5. They are available of course from the authors upon request.

5 we report the tail areas of the *F*-tests performed in testing for fiscal policy asymmetries. In this specification with contemporaneous and four lagged fiscal shocks we are able to perform the following tests: First, we test the null hypothesis that the coefficient of a contemporaneous positive government spending shock is equal to the coefficient of a contemporaneous negative government spending shock ($H_0: \beta_{10} = \beta_{20}$) or in other words that an expansionary government spending shock has a symmetric effect on output as an equal contractionary government spending shock. In a similar manner, we test for symmetric effects of the expansionary and contractionary government revenue shocks with the null hypothesis ($H_0: \beta_{30} = \beta_{40}$). Next, we test whether equivalent in terms of their impact on government deficit fiscal policies have asymmetric effects on the growth rate of real private GNP and its cyclical component. First, policies that increase the deficit, i.e. positive government spending and negative government revenue shocks ($H_0: \beta_{10} = -\beta_{40}$) and second, shocks that lead to fiscal consolidation, a decrease in government spending and an increase in government revenue ($H_0: \beta_{20} = -\beta_{30}$).

Moreover, we perform *F*-tests that all lagged coefficients of the unanticipated fiscal policy shocks are jointly equal to zero: $H_0: \beta_{j0} = \beta_{j1} = \beta_{j2} = \beta_{j3} = \beta_{j4} = 0$, $j = 1,...,4$. Finally, in the last two rows of Table 5 we test for asymmetric cumulative effects of contractionary and expansionary unanticipated fiscal policy shocks:

$$H_0: \beta_{j0} + \beta_{j1} + \beta_{j2} + \beta_{j3} + \beta_{j4} = \beta_{k1\,0} + \beta_{k\,1} + \beta_{k\,2} + \beta_{k\,3} + \beta_{k\,4},$$
$$j = 1, 3 \text{ and } k = 2, 4$$

Table 5 summarizes the regressions results and the hypotheses testing evidence. According to these, the estimated coefficients of all unanticipated fiscal policy shocks that are statistically significant appear to have the correct sign. In System 2 where the dependent variable is the growth rate of real private GNP, no shock appears statistically significant. In System 4 with the same dependent variable but shocks identified from a VAR using the Divisia MZM as the exogenous monetary aggregate variable only the contemporaneous unexpected negative government spending shock appears statistically significant with a coefficient of 0.310 and a *p*-value of 0.075. In System 6 where the dependent variable is now the cyclical component of real private GNP, three types of unanticipated government shocks are significant: the contemporaneous positive and negative government spending with coefficients 0.237 and 0.423 respectively and the second lag of the positive government revenue shock with a coefficient of -0.161. the *p*-values of these estimates are 0.095, 0.012 and 0.031 respectively. In system ML, three types of unanticipated government shocks are significant: the third lag of the positive and the negative government spending shock and the third lag of the negative government revenue shock. Only the third lag of the positive government shock has the expected sign. The other two shocks have the opposite sign as expected from the theory. From these latter results it seems that both positive and negative government spending shocks have a positive impact, after three quarters, on the private real GNP. On the other hand, an unexpected reduction in government revenues will has a negative impact on the real private GNP in a three quarters period.

In the lower part of Table 5 we report the results of the *F*-tests discussed above. We find evidence of unanticipated fiscal shock asymmetries in three cases in System 6 and in system ML. In system 6, where the dependent variable is the cyclical component of real private GNP the three reported asymmetries are: a) asymmetry in the type of contractionary policy as negative government spending shocks appear to have a significant impact on the cycle of real private GNP while positive government revenue shocks appear statistically insignificant and the *p*-value of the test of equality between the two is 0.038; b) the joint test that all contemporaneous and lagged negative unanticipated government spending shocks appears statistically significant with *p*-value 0.094 while the positive joint government spending shock appears statistically insignificant; c) the joint test that all contemporaneous and lagged positive unanticipated government revenue shocks appears statistically significant with *p*-value 0.031 while the negative joint government revenue shock appears statistically insignificant. According to these results, only a contractionary unanticipated fiscal policy will have an impact on the cyclical component of GNP either through decreased government spending or increased government revenue. The joint impact of expansionary fiscal shocks appears statistically insignificant. In the last four rows of Table 5 we perform *F*-tests of equality between the cumulative effects of unanticipated fiscal policy shocks. We cannot reject any of these hypotheses.

In system ML, in which the dependent variable is the growth rate of the real private GNP there is an asymmetry in the expansionary fiscal policy, where the positive government spending shock has a greater impact on GNP than the negative government revenue shock. Moreover, the test that the impact of all contemporaneous and lagged negative unanticipated government revenue shocks is statistically different than the impact of all contemporaneous and lagged positive unanticipated government revenue shocks.

## 6. Conclusions

The aim of this paper was to empirically test the effects of fiscal policy shocks on the level and growth rate of real output and reveal possible asymmetries in fiscal policy implementation. The data are quarterly over the period 1967:1 to 2011:4. In doing so, we used two alternative vector autoregressive systems as well as a support vector regression model in order to construct the fiscal policy shocks. These systems differ in the monetary aggregate used as one of the exogenous variables: a simple sum MZM and a CFS Divisia MZM. From each one of these systems we extracted four types of shocks: a negative and a positive government spending shock and a negative and a positive government revenue shock. These eight sets of unanticipated fiscal shocks were used next to empirically examine their effects on the level and growth rate of real private GNP in two sets of regressions: one that assumes only contemporaneous effects of the shocks on output and one that is augmented with four lags of each fiscal shock. Our results are summarized as follows:

In the regressions with no lagged shocks we detect some asymmetries across the three systems that extracted errors from the VARs used, with respect to fiscal policy shocks. When the dependent variable is the growth rate of real private GNP expansionary and contractionary unanticipated fiscal policies have asymmetric effects: the expansionary fiscal policy through either a positive unanticipated government spending shock (SGP) or a negative unanticipated government revenue shock (SRN) is statistically significant with *p*-values of 0.008 and 0.052 respectively while a contractionary fiscal policy through either a negative unanticipated government spending shock (SGN) or a positive unanticipated government revenue shock (SRP) is not significant. When the dependent variable is the cyclical component of real private GNP it appears that only a positive unanticipated government revenue shock has some impact on the cycle. We find that in general, expansionary unanticipated fiscal policy shocks have asymmetric effects and appear to affect real private output more than contractionary fiscal shocks as the later appear insignificant. Expansionary fiscal policy is significant either through spending or revenue even though the coefficient of the positive government spending shock is more than three times larger than the coefficient of the negative government revenue shock in the systems with growth rate of GNP as the dependent variable and more than two times larger in the when the dependent variable is the GNP cyclical component.

Finally, in the systems with lags, only the contemporaneous unexpected negative government spending shock appears statistically significant in System 4. In System 6, where the dependent variable is now the cyclical component of real private GNP, three types of unanticipated shocks are significant: the contemporaneous positive and negative government spending and the second lag of the positive government revenue shock. In system ML, the third lags of the unanticipated positive and negative government spending shocks and the third lag od the negative government spending shock are statistically significant, whereas only the positive government spending shock has the sign as it is expected by the theory.

In these systems, we find evidence of asymmetries in five cases and when the dependent variable is the cyclical component of real private GNP or the growth rate of the real private GNP: a) asymmetry in the type of contractionary policy (negative government spending shocks have a significant impact on the cycle of real private GNP while positive government revenue shocks appear statistically insignificant) b) the joint test that all contemporaneous and lagged negative unanticipated government spending shocks appears statistically significant while the positive joint government spending shock appears statistically insignificant c) the joint test that all contemporaneous and lagged positive unanticipated government revenue shocks appears statistically significant while the negative joint government revenue shocks appear statistically insignificant d) the positive government spending shock has a greater impact on GNP than the negative government revenue shock and e) the impact of all contemporaneous and lagged negative unanticipated government revenue shocks is statistically different than the impact of all contemporaneous and lagged positive unanticipated government revenue shocks.

Finally, the use of the support vector regression method did not gave to our results any extra explanatory power than the classical Blanchard and Perotti method.

## References

Alesina, A., Ardagna, S., 2009. Large changes in Fiscal policy: Taxes versus Spending. *NBER working Paper series*. 15348.

Alesina, A., Perotti, R., 1997. The welfare state and competitiveness. *American Economic Review*, 87 (5). Pp. 921-939

Barro, R., 1974. Are government Bonds net wealth? *Journal of Political Economy*, 82 (6). Pp. 1095-1117.

Barro, R. and Redlick, C. 2009. Macroeconomic effects from government purchases and taxes. *NBER* Working Paper 15369.

Baum A., and Koester G. 2011. The impact of fiscal policy on economic activity over the business cycle – evidence from a threshold VAR analysis. *Deutsche Bank,* Discussion Paper, series1: economic studies, No 03/2011.

Bertola, G, Drazen, A., 1993. Trigger points and budget cuts: explaining the effects of fiscal austerity. *American Economic Review*, Vol. 83, pp. 11-26.

Blanchard, O., 1990. Suggestions for a new set of fiscal indicators. *OECD Economics Department, Working Paper.* 79.

Blanchard, O., Perotti, R., 2002. An Empirical Characterization of the Dynamic Effects of Changes in Government Spending and Taxes on Output. *Quarterly Journal of Economics*, 117 (4), 1329–68.

Cogan, J., Cwik, T., Taylor, J., Wieland, V. 2009. New Keynesian versus old kenesian government spending multipliers. *NBER* Working Paper 14782.

Cortes C. and Vapnik V. (1995), Support-Vector Networks, Machine Learning, vol 20, pp. 273-297.

Gogas Periklis, Papadimitriou Theophilos and Takli Elvira (2013) Comparison of Simple Sum and Divisia monetary aggregates in GDP forecasting: a Supprot Vector Machines approach", Economics Bulletin, vol. 33(2), pages 1101-1115.

Cover, J. 1992. Asymmetric Effects of Positive and Negative Money Supply Shocks. *Quarterly Journal of Economics*, 107(4), 1261- 1282.

Gali, J., Lopez-Salido, J. D. and Valles, J. 2007. Understanding the Effects of Government Spending on Consumption. *Journal of the European Economic Association* 5(1), pp. 227-270.

Giavazzi, F. and Pagano, M., 1990. Can severe fiscal contractions be expansionary? Tales of two small European countries. *NBER, Macro- Economics Annual* 5, pp. 75-111.

Kollias, C. and Palaiologou, S. 2006. Fiscal Policy in the European Union: Tax and spend, spend and tax, Fiscal Synchronization or Institutional separation?, *Journal of Economic Studies,* 33(2), pp.108-120.

Leeper E., Walker T., and Shu-Chun Susan Yang, 2010. Government investment and fiscal stimulus in the short and long runs. *Journal of Monetary Economics* 57: 1000-1012.

Mountford, A., and Uhlig, H., 2009. What are the effects of fiscal policy shocks? *Journal of Applied Econometrics*, 24: 960-992.

Owoye, O. 1995. The causal relationship between taxes and expenditures in the G7 countries: cointegration and error correction models. *Applied Economics Letters*, Vol. 2, pp. 19-22.

Papadimitriou Theophilos, Gogas Periklis, Vasilios Plakandaras and Mourmouris John (2013), "Forecasting the insolvency of US banks with Support Vector Machines, Journal of Computational Economics and Econometrics, vol. 3 no 1/2.

Payne, J.E. 2003. A survey of the international empirical evidence on the tax-spend debate. *Public Finance Review*, Vol. 31, pp. 302-24

Pereira, M. and Lopes, A. 2010. Time varying fiscal policy in U.S. *Banko de Portugal,* Working Paper 21/2010.15

Plakandaras Vasilios, Papadimitriou Theophilos,and Gogas Periklis (2013), Directional Forecasting in financial time series using Support Vector machines: the USD/ Euro exchange rate, Journal of Computational Optimization in Economics and Finance, vol. 5(2).

Romer, C, RomerD. 2010. The macroeconomic effects of tax changes: Estimates based on new measure of fiscal shocks. *American Economic Review* 100: 763-801.

Sims, C. A., 1980. Macroeconomics and Reality. *Econometrica*, 48, 1–48.

Sims, C. A., Stock, J. and Watson W., M. 1990. Inference in Linear Time Series models with some unit roots. *Econometrica,* 58, 114-44.

Tagkalakis A., 2008. The effects of fiscal policy on consumption in recessions and expansions. *Journal of Public Economics* 92(5-6):1486-1508.

Vapnik, V., Boser, B. and Guyon, I. 1992. A training algorithm for optimal margin classifiers. Fifth Annual Workshop on Computational Learning Theory, Pittsburgh, ACM, pp.144–152.

**Table 1.** Unit Root Tests

| Variable | A. ADF Test | | | B. KPSS Test | | | C. Elliott et al. Test | | | Decision |
|---|---|---|---|---|---|---|---|---|---|---|
| | Level | 1st Diff. | | Level | 1st Diff. | | Level | 1st Diff. | | |
| | Null Hypothesis: I(1) | | | Null Hypothesis: I(0) | | | Null Hypothesis: I(1) | | | |
| | Probability margin | | | LM-Stat | | | Test statistic | | | |

Endogenous Variables: Real Variables

| | | | | | | | | | | |
|---|---|---|---|---|---|---|---|---|---|---|
| $r$ | 0.799 | 0.000 | *** | 0.211 | ** | 0.036 | 20.609 | 1.635 | *** | I(1) |
| $g$ | 0.745 | 0.000 | *** | 0.122 | * | 0.081 | 24.006 | 1.336 | *** | I(1) |
| $y$ | 0.114 | 0.000 | *** | 0.113 | | 0.044 | 4.350 * | 1.352 | *** | I(1) |

Exogenous Variables: Monetary Variables

Simple Sum
| | | | | | | | | | | |
|---|---|---|---|---|---|---|---|---|---|---|
| MZM | 0.856 | 0.000 | *** | 0.205 | ** | 0.114 | 14.708 | 1.331 | *** | I(1) |

CFS Divisia
| | | | | | | | | | | |
|---|---|---|---|---|---|---|---|---|---|---|
| MZM | 0.038 ** | 0.014 | ** | 0.087 | | 0.060 | 1.473 *** | 3.445 | *** | I(1) |
| TB3 | 0.323 | 0.000 | *** | 0.224 | *** | 0.028 | 13.186 | 1.260 | *** | I(1) |

The tests are done with an intercept and a trend.
*, ** or ***, denote denote a rejection of the null hypothesis at the 10%, 5% or 1% level.
The 10%, 5% and 1% critical values for the KPSS tests are 0.119, 0.146 and 0.216 respectively.
The 10%, 5% and 1% critical values for the Elliot-Rothenberg-Stock tests are 6.845, 5.656 and 4.094 respectively.

**Table 2**. Systems Employed in Fiscal Policy Shock Extraction

| | Dependent Variable | | | Exogenous Variables | | | Lags |
|---|---|---|---|---|---|---|---|
| | r | g | y | TB3 | Simple Sum MZM | Divisia MZM | |
| VAR 1 | ✓ | ✓ | ✓ | ✓ | ✓ | | 4 |
| VAR 2 | ✓ | ✓ | ✓ | ✓ | | ✓ | 4 |

**Table 3.** Johansen Maximum Likelihood Cointegration Tests

| Endogenous | Exogenous | VAR Lags | Normality J-B joint test | Serial Correlation LM test | Null Hypothesis | $\lambda_{trace}$ | | $\lambda_{max}$ | | Coint. Vectors |
|---|---|---|---|---|---|---|---|---|---|---|
| r, g, y | TB3 Sum MZM | 4 | 0.000 | 0.240 | $r = 0$ | 0.008 | *** | 0.025 | ** | 1 |
| | | | | | $r <= 1$ | 0.122 | | 0.305 | | |
| | | | | | $r <= 2$ | 0.045 | ** | 0.045 | ** | |
| r, g, y | TB3 Divisia MZM | 4 | 0.000 | 0.288 | $r = 0$ | 0.001 | *** | 0.013 | ** | 1 |
| | | | | | $r <= 1$ | 0.019 | ** | 0.215 | | |
| | | | | | $r <= 2$ | 0.004 | *** | 0.004 | *** | |

One, two and three asteriscs denote rejection of the null hypotheis at the 10%, 5% and 1% levels respectively.

**Table 4.** Fiscal Policy Shocks on the Level of Real Private GNP

| | System 1 | | | System 3 | | | System 5 | | | System ML | | |
|---|---|---|---|---|---|---|---|---|---|---|---|---|
| | Coefficient | p-value | | Coefficient | p-value | | Coef. | prob. | | Coef. | prob. | |
| C | 0.002 | 0.219 | | 0.002 | 0.266 | | -0.001 | 0.303 | | 0.000 | 0.8425 | |
| d(y(-1)) | 0.214 | 0.004 | *** | 0.206 | 0.020 | ** | 1.101 | 0.000 | *** | 0.191 | 0.0284 | ** |
| d(y(-2)) | 0.155 | 0.027 | ** | 0.136 | 0.030 | ** | -0.274 | 0.003 | *** | 0.131 | 0.0634 | * |
| d(TB3) | 0.002 | 0.004 | *** | 0.002 | 0.058 | * | 0.003 | 0.026 | ** | 0.002 | 0.0375 | ** |
| d(MZM) | 0.118 | 0.002 | *** | 0.197 | 0.001 | *** | 0.080 | 0.172 | | 0.205 | 0.0026 | *** |
| SGP | 0.406 | 0.008 | *** | 0.364 | 0.077 | * | 0.138 | 0.412 | | 179.82 | 0.3329 | |
| SGN | 0.217 | 0.113 | | 0.226 | 0.226 | | 0.436 | 0.012 | | -68.69 | 0.7095 | |
| SRP | -0.102 | 0.166 | | -0.123 | 0.194 | | -0.124 | 0.061 | * | 17.76 | 0.6951 | |
| SRN | -0.124 | 0.052 | * | -0.119 | 0.087 | * | -0.067 | 0.295 | | 13.028 | 0.7319 | |
| | | | | | | | | | | | | |
| F-Tests | | | | | | | | | | | | |
| SGP=SGN | | 0.444 | | | 0.690 | | | 0.313 | | | - | |
| SRP=SRN | | 0.848 | | | 0.979 | | | 0.566 | | | - | |
| SGP=SRN | | 0.088 | * | | 0.256 | | | 0.677 | | | - | |
| SGN=SRP | | 0.460 | | | 0.578 | | | 0.109 | | | - | |

Note: *, ** and ***, denote a rejection of the null hypothesis at the 0.10, 0.05 and 0.01 levels respectively

**Table 5.** Fiscal Policy Shocks on Real Private GNP using Four Lags

|  | System 2 | | | System 4 | | | System 6 | | | System ML | | |
|---|---|---|---|---|---|---|---|---|---|---|---|---|
|  | Coef. | prob. | | Coef. | prob. | | Coef. | prob. | | Coef. | prob. | |
| C | 0.000 | 0.953 | | 0.001 | 0.668 | | -0.003 | 0.124 | | 0.003 | 0.65 | |
| d(y((-1)) | 0.265 | 0.003 | *** | 0.240 | 0.010 | ** | 0.943 | 0.000 | *** | 0.162 | 0.06 | * |
| d(y((-2)) | 0.203 | 0.019 | ** | 0.195 | 0.032 | ** | 0.020 | 0.873 | | 0.168 | 0.03 | ** |
| d(y((-3)) | 0.050 | 0.577 | | 0.029 | 0.730 | | -0.146 | 0.132 | | -0.106 | 0.13 | |
| d(y((-4)) | -0.008 | 0.922 | | -0.016 | 0.845 | | -0.045 | 0.560 | | -0.005 | 0.94 | |
| d(TB3) | 0.001 | 0.428 | | 0.001 | 0.272 | | 0.002 | 0.075 | * | 0.001 | 0.17 | |
| d(TB3(-1)) | 0.001 | 0.420 | | 0.002 | 0.077 | * | 0.003 | 0.017 | ** | -0.000 | 0.52 | |
| d(TB3(-2)) | 0.000 | 0.957 | | 0.001 | 0.551 | | 0.002 | 0.028 | ** | 0.001 | 0.40 | |
| d(TB3(-3)) | 0.001 | 0.725 | | 0.000 | 0.800 | | 0.003 | 0.017 | ** | 0.000 | 0.92 | |
| d(TB3(-4)) | -0.003 | 0.010 | ** | -0.003 | 0.000 | *** | -0.001 | 0.235 | | -0.002 | 0.00 | ** |
| d(mzm) | 0.081 | 0.138 | | 0.232 | 0.032 | ** | 0.123 | 0.305 | | 0.099 | 0.24 | |
| d(mzm(-1)) | 0.126 | 0.032 | ** | 0.233 | 0.032 | ** | 0.228 | 0.011 | ** | 0.175 | 0.05 | * |
| d(mzm(-2)) | -0.012 | 0.858 | | -0.035 | 0.727 | | 0.036 | 0.695 | | 0.071 | 0.42 | |
| d(mzm(-3)) | -0.039 | 0.662 | | -0.199 | 0.074 | * | -0.093 | 0.286 | | -0.193 | 0.01 | ** |
| d(mzm(-4)) | -0.004 | 0.937 | | -0.007 | 0.917 | | -0.080 | 0.164 | | 0.004 | 0.94 | |
| SGP | 0.296 | 0.214 | | 0.275 | 0.210 | | 0.237 | 0.095 | * | -67.38 | 0.73 | |
| SGP(-1) | 0.015 | 0.921 | | 0.077 | 0.633 | | 0.052 | 0.709 | | 135.88 | 0.35 | |
| SGP(-2) | -0.120 | 0.370 | | -0.044 | 0.761 | | -0.166 | 0.156 | | -90.03 | 0.55 | |
| SGP(-3) | -0.257 | 0.151 | | -0.213 | 0.199 | | -0.136 | 0.290 | | 475.36 | 0.01 | ** |
| SGP(-4) | -0.115 | 0.431 | | -0.173 | 0.262 | | -0.118 | 0.326 | | -233.30 | 0.23 | |
| SGN | 0.273 | 0.165 | | 0.310 | 0.075 | * | 0.423 | 0.012 | ** | 81.66 | 0.68 | |
| SGN(-1) | -0.180 | 0.283 | | -0.131 | 0.419 | | -0.149 | 0.283 | | 100.56 | 0.48 | |
| SGN(-2) | -0.052 | 0.690 | | -0.063 | 0.606 | | 0.043 | 0.754 | | 48.03 | 0.77 | |
| SGN(-3) | -0.204 | 0.184 | | -0.193 | 0.204 | | 0.046 | 0.705 | | -515.04 | 0.01 | ** |
| SGN(-4) | 0.118 | 0.425 | | 0.145 | 0.334 | | 0.222 | 0.173 | | 216.79 | 0.29 | |

**Table 5** (continued). Fiscal Policy Shocks on Real Private GNP using Four Lags

| | | | | | | | | | |
|---|---|---|---|---|---|---|---|---|---|
| SRP | -0.073 | 0.461 | -0.097 | 0.345 | -0.064 | 0.273 | | 20.0 | 0.61 |
| SRP(-1) | 0.135 | 0.123 | 0.116 | 0.167 | -0.065 | 0.344 | | 26.2 | 0.35 |
| SRP(-2) | 0.076 | 0.373 | 0.057 | 0.493 | -0.161 | 0.031 | ** | -4.32 | 0.90 |
| SRP(-3) | -0.054 | 0.420 | -0.048 | 0.459 | -0.069 | 0.168 | | 26.11 | 0.48 |
| SRP(-4) | 0.056 | 0.455 | 0.057 | 0.416 | 0.075 | 0.308 | | -12.77 | 0.73 |
| SRN | -0.161 | 0.071 | -0.187 | 0.011 | -0.206 | 0.001 | | -5.66 | 0.87 |
| SRN(-1) | -0.030 | 0.710 | -0.024 | 0.738 | 0.045 | 0.610 | | -25.98 | 0.37 |
| SRN(-2) | 0.017 | 0.788 | 0.058 | 0.309 | 0.112 | 0.070 | | 5.84 | 0.88 |
| SRN(-3) | 0.049 | 0.464 | 0.064 | 0.315 | 0.070 | 0.214 | | 104.03 | 0.00 | *** |
| SRN(-4) | 0.020 | 0.721 | 0.020 | 0.724 | -0.015 | 0.723 | | 18.34 | 0.55 |

| | System 2 | System 4 | System 6 | | System ML | |
|---|---|---|---|---|---|---|
| F-Tests | | | | | | |
| SGP=SGN | 0.951 | 0.923 | 0.478 | | 0.652 | |
| SRP=SRN | 0.578 | 0.530 | 0.118 | | - | |
| SGP=SRN | 0.590 | 0.701 | 0.840 | | 0.005 | *** |
| SGN=SRP | 0.293 | 0.227 | 0.038 | ** | - | |
| joint SGP=0 | 0.686 | 0.843 | 0.638 | | 0.632 | |
| joint SGN=0 | 0.896 | 0.807 | 0.094 | * | 0.855 | |
| joint SRP=0 | 0.376 | 0.585 | 0.031 | ** | 0.507 | |
| joint SRN=0 | 0.574 | 0.656 | 0.968 | | 0.289 | |
| Σ SGP = Σ SGN | 0.841 | 0.799 | 0.179 | | 0.441 | |
| Σ SRP = Σ SRN | 0.348 | 0.495 | 0.161 | | 0.000 | *** |
| Σ SGP = Σ SRN | 0.628 | 0.768 | 0.676 | | 0.509 | |
| Σ SGN = Σ SRP | 0.748 | 0.561 | 0.437 | | 0.975 | |

Note: *, ** and ***, denote a rejection of the null hypothesis at the 0.10, 0.05 and 0.01 levels respectively